\RequirePackage{fix-cm}
\documentclass[smallextended]{svjour3}       
\smartqed  
\usepackage[utf8]{inputenc}
\usepackage[english]{babel}
\usepackage{amsmath}
\usepackage{mathtools}
\usepackage{amsfonts}
\usepackage{bbold}
\usepackage{subcaption}
\usepackage{amssymb}
\usepackage{graphicx}
\usepackage{subcaption}
\usepackage{natbib}
\usepackage{sidecap}
\usepackage[citecolor=blue,]{hyperref}
\usepackage{lscape}
\usepackage{numprint} 
\npdecimalsign{.}
\usepackage{float}
\usepackage[shortlabels]{enumitem}
\begin{document}

\title{Multivariate Distributional Stochastic Frontier Models
}


\author{Rouven Schmidt         \and
        Thomas Kneib 
}


\institute{Rouven Schmidt \at
              TU-Clausthal \\
              Tel.: +49 (0) 551 - 39 25431\\ 
              \email{rouven.schmidt@tu-clausthal.de}           
           \and
           Thomas Kneib \at
              Georg-August University G\"ottingen \\
              Tel.: +49 (0) 551 / 39 25678\\ 
              \email{tkneib@uni-goettingen.de}           
}

\date{Received: date / Accepted: date}

\maketitle


\begin{abstract}
The primary objective of Stochastic Frontier (SF) Analysis is the deconvolution of the estimated composed error terms into noise and inefficiency. Assuming a parametric production function (e.g. Cobb-Douglas, Translog, etc.), might lead to false inefficiency estimates. To overcome this limiting assumption , the production function can be modelled utilizing P-splines.  Application of this powerful and flexible tool enables modelling of a wide range of production functions. Additionally, one can allow  the parameters of the composed error distribution to depend on covariates in a functional form. The SF model can then be cast into the framework of a \textit{Generalized Additive Model for Location, Scale and Shape} (\textit{GAMLSS}).  
Furthermore, a decision-making unit (DMU) typically produces multiple outputs. It does this by operating several sub-DMUs, which each employ a production process to produce a single output. Therefore, the production processes of the sub-DMUs are typically not independent. Consequently, the inefficiencies  may be expected to be dependent, too. In this paper, the \textit{Distributional Stochastic Frontier Model} (\textit{DSFM}) is introduced.  The multivariate distribution of the composed error term is modeled using a copula. As a result, the presented model is a generalization of the model for seemingly unrelated stochastic frontier regressions by Lai and Huang (2013).
\end{abstract}



\section{Introduction}
Stochastic Frontier (SF) analysis has been widely used in  productivity and efficiency studies to describe and estimate production and cost frontier models. In the parametric setting, such analysis requires that the unknown functional form of the production or cost function has to be specified a priori to the analysis, e.g. Cobb-Douglas or Translog functions. SF models by \cite{aigner1977formulation} and \cite{meeusen1977efficiency} allocate the difference between the production (cost) function and observed output to the composed error, which consists of a measurement error and an (in)efficiency term. 
Assuming a false production (cost) function thus leads to false (in)efficiencies, results in fallacious conclusions. \\
To overcome this assumption, several semi- and nonparametric approaches have been introduced in the SF literature. The approach by \cite{fan1996semiparametric} estimates the production (cost) function in the first step based on a kernel density estimation of the conditional mean function. In the second step, the (in)efficiency parameters are then estimated via pseudolikelihood. While \cite{kumbhakar2007nonparametric} utilizes a  local maximum likelihood technique, the StoNED approach by \cite{kuosmanen2008representation} and \cite{kuosmanen2010data} applies piecewiese linear function to approximate the regression function, which minimizes the $L_2$-norm. The model by \cite{ferrara2017semiparametric} extends the work of \cite{fan1996semiparametric} by utilizing the GAMLSS framework in the first step of the estimation. \cite{klein2020modelling} introduced a semiparametric flexible structured approach to stochastic frontier analysis in panel models in the Bayesian setting. They showed that the SF model can be written as distributional regression case, thus it can directly written as a \textit{GAMLSS}, introduced by \cite{stasinopoulos2007generalized}. Thus, only one estimation step is required. Any functional form of the production or cost function can be approximated, and the introduced \textit{GAMLSS} is able to deal with heterogeneity coming from the inefficiency as well as the measurement error. Shape constraints, coming from economic theory, on the production or cost function can be enforced utilizing the ideas from \cite{pya2015shape}. \\
Further, such SF analysis typically assumes that a DMU employs a single production process or technology to produce a single output using multiple inputs. To analyze DMUs with multiple outputs, one assumes that a DMU may have several sub-DMUs. Each produces one output and has its own set of inputs. Since the sub-DMUs belong to the same DMU, they may be subject to the same random shocks as the parent DMU. 
Thus, the sub-DMUs technical (in)efficiencies terms may not be independent. Under this circumstance, a system of multiple stochastic frontier regressions on the sub-DMUs is a more appropriate representation of a DMU’s operation and performance. \cite{lai2013maximum} proposed the copula-based maximum likelihood (ML) approach to estimate the stochastic frontier models with correlated composite errors. The recent work by \cite{lai2020maximum} extends the multiple output model to account for panel data facilitating a simulated ML approach. Thus, the researcher can take explicit account of sub-DMU-specific heterogeneity and inefficiency. Consequently, allowing for the detection and measurement of effects which cannot be observed in cross-sectional data.  This work proposes a fast, accurate and reliable maximum likelihood approach to estimate the multiple output panel data model utilizing smooth terms which is inherently more efficient than \cite{lai2020maximum}. The hereby introduced \textit{Distributional Stochastic Frontier Model} (\textit{DSFM}) allows for are more complex structure of the dependence between the outputs. In summary, the \textit{DSFM} is an extension of the SF model which allows modelling the influence of covariates beyond the mean in a linear and non-linear way. Additionally, the DMU specific random effect and  the  dependence between severall sub-DMUs can be captured.
Further, it is shown that the \textit{DSFM} can be written as a special case of the \textit{Generalized Joint Regression Model} (\textit{GJRM}) introduced by \cite{marra2017bivariate}.  In summary, the main aim of this paper is threefold:
\begin{enumerate}[(I)]
    \item Cast the SF model in the framework of distributional regression models
    \item Show that the shape-constraint spline approach of \cite{pya2015shape} can be applied to distributional regression models
    \item Generalize the model to allow for multiple outputs and estimation via penalized ML
\end{enumerate}
The paper is structured as follows. In Section \ref{model} the \textit{GJSFM} is formulated. Estimation of the model is described in detail in Sect. \ref{estimation}.
The performance of the suggested approach is then investigated in a Monte Carlo simulation study in Sect. \ref{MC simulation}.
In Sect. \ref{conclusions} the authors conclude.

\section{Distributional Stochastic Frontier Model} \label{model}
Within the following section, the details of the \textit{DSFA} are presented. First, it is shown that the model for a single sub-DMU can be written as a \textit{GAMLSS}. Thereafter, details on the inclusion of linear, non-linear and random effects are provided. In the third part, the concept is generalized to account for multiple sub-DMUs utilizing copulas.

\subsection{Model Specification}
Suppose a DMU $n \in \{1,2, \hdots, N\}$ produces $M$ outputs at time $t_n \in \{1,2,\hdots,T_n \}$ with $N, M, T_n \in \mathbb{N} \setminus\{0\}$. Each output $m \in \{ 1,2,\ldots, M \}$ is produced by a sub-DMU under a  production technology. The production or cost frontier is represented as an SF model
\begin{align}
\label{eq1}
Y_{nt_nm} &= \underbrace{\eta^{\mu}(\boldsymbol{x}^{\mu}_{nt_nm})}_{\mu_{nt_nm}} + \underbrace{V_{nt_nm}  + s \cdot U_{nt_nm}}_{\mathcal{E}_{nt_nm}} \qquad.
\end{align}
For $s=-1$, $\eta^\mu(\cdot)$ is the production function and $x^\mu_{nt_nm}$ are the log inputs for DMU $n$ sub-DMU $m$ at time $t_n$. Alternatively, if $s=1$, $\eta^\mu(\cdot)$ is the cost function and $x_{nt_nm}^\mu$ are the log costs. The function $\eta^\mu(\cdot)$  defines a production (cost) relationship between inputs and the output. Assuming that $\eta^\mu(\cdot)$ belongs to a parametric family of functions is quite restrictive and can lead to a serious modelling bias resulting in misleading conclusions about the link between inputs and output, as shown by \cite{giannakas2003choice}. 
Instead, a more flexible approach utilizing P-splines is chosen, similar to \cite{klein2020modelling}. Here, $\eta^\mu(\cdot)$ can be any theoretically valid production function. Optionally, one can enforce that it satisfies the regularity conditions by \cite{chambers1988applied} utilizing Shape Constrained P-Splines (SC-Splines) introduced by \cite{pya2015shape}. The concept is visualized in Figure \ref{fig:prodfun}:
\begin{figure}[H] 
         \centering
         \includegraphics[width=0.75\textwidth]{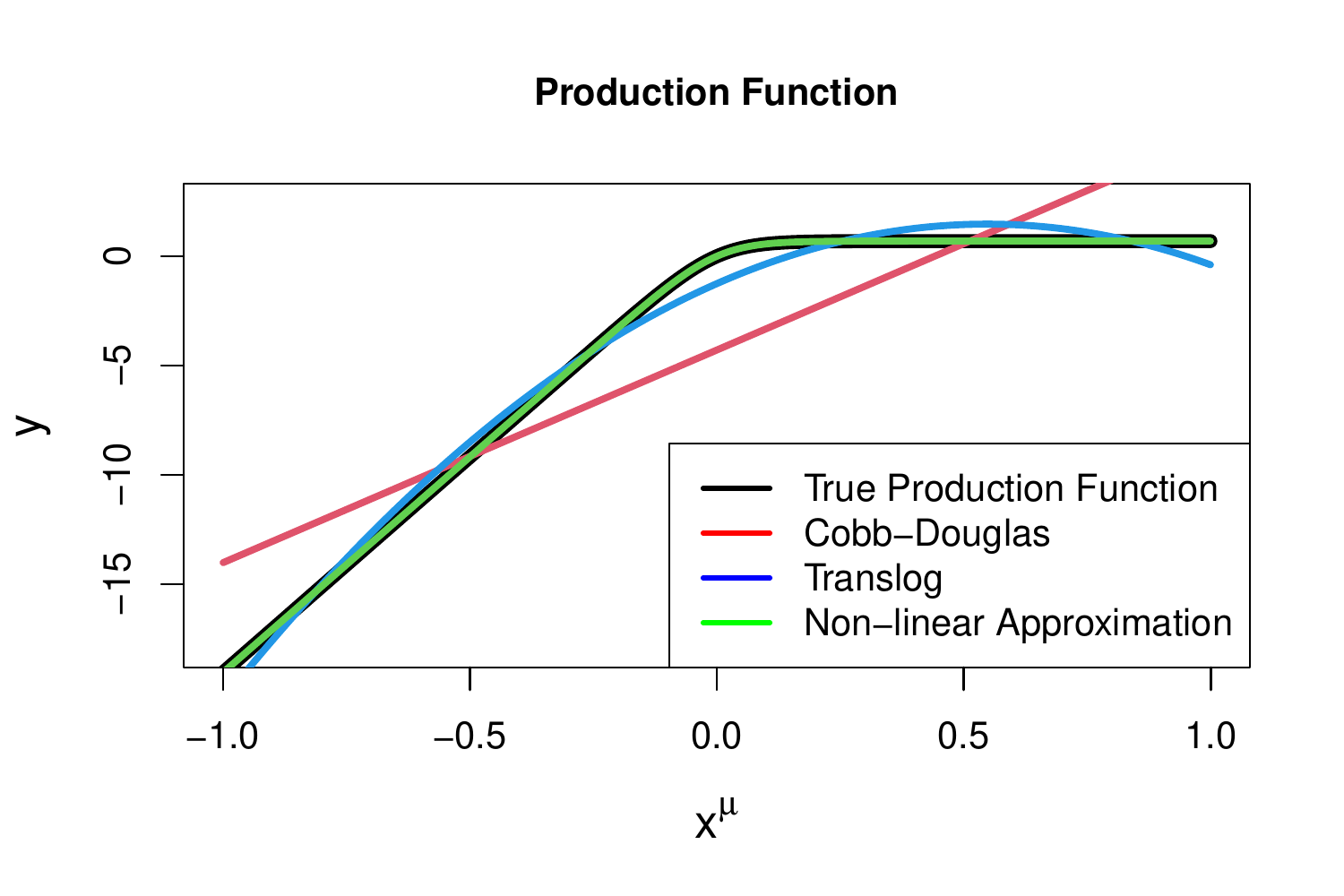}
        \caption{The log of the true production function $1+\tanh(\log(x^\mu_{nt_nm}) \cdot \pi^2)$ is plotted in black which fulfills the regularity conditions of \cite{chambers1988applied} with $\exp(x^\mu_{nt_nm}) \in [\exp(-1), \exp(1)]$. Even though there is no measurement error or inefficiency, the fit of the Cobb-Douglas and Translog Production function are poor. They would only allow  for unbiased estimates of the inefficiency in two and four points, respectively. Thus, the primary goal of SF Analysis could not be achieved. The Non-linear approximation fits the function perfectly. Here, B-splines were utilized.}
        \label{fig:prodfun}
\end{figure}
The unobserved measurement error and the (in)efficiency are represented via the random variables $V_{nt_nm}$ and $U_{nt_nm}$. The parameters of their corresponding distributions are denoted as $\sigma_{Vnt_nm}$ and $\sigma_{Unt_nm}$ \footnote{If the distributions have more than one parameter, the notation is $\sigma_{V1nt_nm},\sigma_{V2nt_nm},\hdots, \sigma_{VP_Vnt_nm}$ and $\sigma_{U1nt_nm},\sigma_{U2nt_nm},\hdots, \sigma_{UP_Unt_nm}$}. The parameters may depend on variables $x^{\sigma_{V}}_{nt_nm}$ and $x^{\sigma_{U}}_{nt_nm}$, e.g. firm size, management skill etc.  The random variables $V_{nt_nm}$ and $U_{nt_nm}$ are assumed to be independent of each other. For $m_1,m_2 \in \{1,\ldots,M\}$ the inefficiency terms of two sub-DMUs $U_{nt_nm_1}$ and $U_{nt_nm_2}$ are not required to be independent, as it is an implicit assumption in disjoint modelling of the sub-DMUs. Thus, allowing the $M$ sub-DMUs of DMU $n$ to share common characteristics, which is reasonable since they are affected by the same management, compliance etc. Furthermore, the random terms  $V_{nt_nm_1}$ and $V_{nt_nm_2}$ may likewise be dependent, allowing a shock to the DMU to potentially influence all sub-DMUs. 
Consequently, the composite errors $\mathcal{E}_{nt_nm_1}$ and $\mathcal{E}_{nt_nm_2}$ are allowed a dependence structure. It is essential to note that within the model it is impossible to decompose the dependence of the composed errors into the dependence of the inefficiencies and the measurement errors. \\
As a result of the distributional assumptions, $Y_{nt_nm} \sim \mathcal{D}(\mu_{nt_nm},\sigma_{Vnt_nm}, \sigma_{Unt_nm})$. \footnote{The corresponding pdf is $f_{Y_{nt_nm}}(y)=\mu_{nt_nm} + \int_{0}^{\infty} f_{V_{nt_nm}}(\epsilon_{nt_nm}-s \cdot u_{nt_nm})f_{U_{nt_nm}}(u_{nt_nm}) du_{nt_nm}$}.
To explain the model, let $\theta$ denote one of the parameters of the distribution, e.g. $\theta_{nt_nm} \in \{\mu_{nt_nm}, \sigma_{Vnt_nm}, \sigma_{Unt_nm} \}$. Let $g_{\theta_k}^{-1}$ be the monotonic response function, which links the additive predictor $\eta^{\theta}(\cdot)$ to the parameter space for the parameter $\theta_{nt_nm}$ via the additive model \footnote{The name distributional stochastic frontier model comes from the fact that all parameters of the distribution are modeled.}:
\begin{align} \label{predictor}
     g_\theta(\theta_{nt_nm}) = \eta^{\theta}(\boldsymbol{x}^{\theta}_{nt_nm})  =& \beta^{\theta}_{m0} + \sum_{j^{\theta}_m= 1}^{J_{m}^{\theta}}  h_{mj^{\theta}_m}^{\theta} (x^{\theta}_{nt_nmj^{\theta}_m}) \qquad.
\end{align}
Thus, the additive predictor  $\eta^{\theta}(\cdot) $is made up by the intercept $\beta_{m0}^\theta$ and $J_m^\theta$ smooths. The smooths can model linear, non-linear and random effects. They are described in the section \ref{Additive Predictor}. The technical efficiency (TE) of DMU $n$ at time $t_n$ for output $m$ can the estimated by the procedures introduced by \cite{jondrow1982estimation} and \cite{battese1988prediction}.

\subsection{Additive Predictor} \label{Additive Predictor}
For simplicity, the index $\theta$ and $m$ are neglected in the following section. In equation (\ref{predictor}), $x_{nt_nj}$ is the $j$-th subvector of the complete covariate vector $\boldsymbol{x}_{n}$.  The vector may contain categorical  or continuous variables. Each function $h_{j} (x_{nt_nj})$  represents an effect, which can be linear, non-linear or random. Regardless of its form, it can be approximated by a linear combination of $Q_j$ basis functions, denoted as $b_{jq_j}(x_{nt_nj})$ and coefficients $\beta_{jq_j}$, thus
\[
h_{j} (x_{nt_nj})=\sum_{q_j=1}^{Q_j} b_{jq_j}(x_{nt_nj}) \beta_{jq_j} \qquad,
\]
which is referred to as a smooth. Consequently, one may write the vector $ ( h_j(x_{1t_1j}), \allowbreak h_j(x_{1t_2j}),  \hdots, h_j(x_{Nt_Nj}) ) $ as the product of design matrix $\boldsymbol{Z_j}[ \sum_{i=1}^{n-1} T_n+t_n, q_j] = b_{jq_j}(x_{nt_nj})$ with the coefficient vector $\boldsymbol{\beta_j}=(\beta_{j1},  \hdots, \beta_{jQ_j})^T$ resulting in
\begin{align} \label{linearpredictor}
      \eta(\boldsymbol{x})  &= \boldsymbol{1}_n \beta_0  + \boldsymbol{Z}_1 \boldsymbol{\beta}_1 + \hdots + \boldsymbol{Z}_J \boldsymbol{\beta}_J =\boldsymbol{Z} \boldsymbol{\beta}
\end{align}
where, $ \boldsymbol{1}_N$ is an $\sum_{i=1}^{N} T_n$-dimensional vector of ones \footnote{Here, $\boldsymbol{Z_j}[i,k]^\theta$ denotes the element of the $i$-th row and $k$-th column of the matrix $\boldsymbol{Z_j}^\theta$ for distribution parameter $\theta$}.
Each $ \boldsymbol{\beta}_j$ is associated with quadratic penalty $\lambda_j \boldsymbol{\beta}_j^T \underbrace{\boldsymbol{P}_j^T \boldsymbol{P}_j}_{\boldsymbol{D}_j} \boldsymbol{\beta}_j$. The penalty enforces specific properties on the $j$-th function, e.g. smoothness. $\boldsymbol{P}_j$ is defined for each smooth as follows.
\begin{itemize}
    \item Linear effects can be modelled  by setting $b_{jq_j}(x_{nt_nj}) = x_{nt_nj}$ with $Q_j=1$. No penalty is assigned to linear effects, e.g. $\boldsymbol{D}_j=\boldsymbol{0}$.
    \item Non-linear effects can be modelled via Penalized B-splines. Here $b_{jq_j}(x_{nt_nj})$ are the known spline basis functions evaluated at $x_{nt_nj}$. The number of knots $Q_j$ as well as the position influences the fit of the function. They may lead to over- or underfitting. To avoid this, \cite{eilers1996flexible} propose to use a relatively  large number of (equidistant) knots and introduce a penalty for functions that are not smooth enough. The smoothness can be approximated by a difference of coefficients of adjacent B-splines:
\[
\boldsymbol{P}_j = \begin{pmatrix}
1 & -2 & 1& & \\
& 1 & -2 & 1& \\
& &  & \ddots& & \\
& & & 1 & -2 & 1 \\
\end{pmatrix} \qquad.
\] 
For more details see \ref{P-splines}.
\item Non-linear effects with shape constraints can be modelled via Shape Constrained Penalized B-splines (SC-splines), 
introduced by \cite{pya2015shape}. From a theoretical economic perspective, the production function is supposed to be monotonically increasing and concave. Consequently, the cost function is monotonically decreasing and convex.
 Rewriting the coeffiecient vector $\boldsymbol{\beta}_j = \boldsymbol{\Sigma}_j \tilde{\boldsymbol{\beta}_j}$ yields the SC-spline, where $\tilde{\boldsymbol{\beta}_j} = (\tilde{ \beta}_{j1}, \tilde{ \beta}_{j2}, \hdots, \tilde{ \beta}_{jQ})^T$. The coefficient vector $\tilde{\boldsymbol{\beta}}_j$ is defined by
 \[
\tilde{ \beta}_{jq_j}=\begin{cases}
 \exp \{ \beta_{jq_j}\} &\text{ if } q_j \in \{q_{j2},q_{j3},\hdots,Q_j\} \\
 \hfill \beta_{jq_j}  &\text{ otherwise  } \qquad.\\
 \end{cases}
 \]
 The matrix $\boldsymbol{\Sigma}$ of dimension $Q_j \times Q_j$ and the quadratic penalty $\boldsymbol{D}_j$ depend solely on the type of constraints, but are independent of the data. For an example, see in the Appendix \ref{SC-splines}.

    \item Random effects are appropriate if the observations are not independent, e.g. a DMU is observed  over several time periods $T_n$. The number of independently observed farms is $Q_j$. The model allows for iid normal random effects with unknown variance. 
The random effects can be written as smooth, as shown by \cite{ruppert2003semiparametric}. Let the basis functions for the smooth $j$ be defined as
\[
b_{jq_j}(x_{nj}) = \begin{cases}
 1 \qquad \text{ if observation $n$ belongs to farm $q_j$ } \\
 0 \qquad \text{ else }
 \end{cases} \quad.
 \]
 The random effect is modeled as the deviation $\beta_{jq_j}$ of farm $q_j$ from the mean effect for all farms. A ridge penalty $\boldsymbol{D}_j = \boldsymbol{I}_N$ is applied. 
\end{itemize}
Further, identifability constraints on smoothers can be enforced by utilizing the approach by \cite{wood2017generalized} which ensures that $\boldsymbol{1}_{\sum_{i=1}^{N} T_n} \boldsymbol{Z} \boldsymbol{\beta} = \boldsymbol{0}_{\sum_{i=1}^{N} T_n}$.  
The overall penalty for $\theta$ is thus $(\boldsymbol{\beta}^\theta)^T \boldsymbol{D}^\theta \boldsymbol{\beta}^\theta$, where $\boldsymbol{D}^\theta = diag(\boldsymbol{0}, \lambda_1^\theta \boldsymbol{D}_1^\theta, \hdots, \lambda_J^\theta \boldsymbol{D}_J^\theta)$.  The smoothing parameter $\lambda_j^\theta \in \mathbb{R}^+$ controls the trade-off between fit and smoothness, and plays a crucial role in determining the shape of $h_{j}^\theta (x^\theta_{nt_nj})$. A large value for $\lambda_j^\theta$ means that the corresponding penalty has a large influence on the parameters of the function during fitting, and vice versa. This penalty must be considered when maximizing the likelihood, which results in the penalized likelihood given by  
\[
ll_{pen}(\boldsymbol{\beta}_m)=ll(\boldsymbol{\beta}_m)- \frac{1}{2}\boldsymbol{\beta}_m^T \boldsymbol{D}_{\lambda m}  \boldsymbol{\beta}_m \qquad,
\]
where  $\boldsymbol{\beta}_m = (\boldsymbol{\beta}^{\mu}_m, \boldsymbol{\beta}^{\sigma_V}_m, \boldsymbol{\beta}^{\sigma_{U}}_m)$ and  $\boldsymbol{D}_{\lambda m}=(\boldsymbol{D}^{\mu}_{\lambda m}, \boldsymbol{D}^{\sigma_V}_{\lambda m}, \boldsymbol{D}^{\sigma_U}_{\lambda m})$. The model parameters $\boldsymbol{\beta}_m$ and $\boldsymbol{D}_{\lambda m}$ can be estimated in the procedure described in \cite{marra2017simultaneous}.

 \subsection{Joint Distribution}
The joint distribution of the $M$ outputs of DMU $n$ is described in the following section.
\cite{dominguez2007matrix} considered a matrix variate closed skew-normal to model the multivariate density $f_{1,2,\ldots,M}(\cdot)$. Unfortunately, it suffers the same shortcomings as the closed-skew normal distribution applied in many SF models. The parameters cannot be estimated via MLE or Method of Moments due to non identifiability. For more details, see \cite{arellano2006unification}. The previously mentioned work provides a solution in the form of the unified-skew normal distribution, which allows only for a linear dependence between the composed error terms. To allow for a more flexible model, a copula approach is chosen to rewrite the multivariate pdf $f_{1,2,\ldots,M}(\cdot)$.
Therefore, the concept of a copula is introduced, in the following.
A copula function $C(\cdot)$ is an $M$-dimensional distribution function with standard uniform margins \[
C(F_{1}(y_{nt_n1}),F_{2}(y_{nt_n2}),\ldots,F_{M}(y_{nt_nM}), \delta_{nt_n}) :
[0, 1]^{M} \rightarrow [0, 1] \qquad.
\]
Here $F_{m}(\cdot)$ denotes the marginal cdf of $y_{nt_nm}$. Further,  $\boldsymbol{\delta}_{nt_n}$ is a vector of parameters of the copula called the dependence parameter, which measures dependence between the marginal cdfs.
A fundamental result of copula theory is Sklar’s theorem, which describes the role that copulas play in the relationship between
multivariate distribution functions and their univariate margins. Sklar’s theorem shows that the univariate margins and the
multivariate dependence can be separated in a such a way that the multivariate dependence structure is represented by the copula
independently of the choice of the margins. Thus, the joint pdf $f_{1,\ldots,M}(\cdot)$ can be written using a copula and the marginal cdfs and pdfs as
\begin{align*}
f_{1,\ldots,M}(y_{nt_n1},y_{nt_n2},\ldots,y_{nM})=&c(F_{1}(y_{nt_n1}),F_{2}(y_{nt_n2}),\hdots,F_{M}(y_{nt_nM}), \delta_{nt_n})\\
&\times\prod_{m=1}^M  f_m (y_{nt_nm})  \qquad,\\
\end{align*}
where $c(\cdot)$ denotes the pdf of the copula\footnote{if it exists, this pdf is defined as $c\left(w_{nt_n1},w_{nt_n2},\ldots,w_{nt_nM} \right)=\frac { \partial C \left(w_{nt_n1},w_{nt_n2},\ldots,w_{nt_nM} \right) } { \partial w_{nt_n1}, \ldots, \partial w_{nt_nM}}$ }.
For the Fisher regularity conditions to hold, only the parametric copula families may be used.
The parameter $\boldsymbol{\delta}_{nt_n}$ can be modelled like the parameters of the marginal distribution, depending on explanatory variables. 
Depending on the copula, the dimension of $\boldsymbol{\delta}_{nt_n}$ changes as well as the monotonic link functions $g_\delta(\cdot)$ which map the additive predictor to the parameter space of the copula parameters. It is important to note, that each copula function enforces certain characteristics regarding the modeled dependence, e.g. the Frank Copula cannot model negative dependence or the Gaussian copula can model only linear dependence. Notably, the independence copula
corresponds to separate regressions for each output $m$. To model dependencies for higher dimensions i.e. $M>2$ with flexible dependence patterns, elliptical or vine copulas may be utilized.

\section{Model Estimation} \label{estimation}
In the following section, the estimation via the penalized Maximum Likelihood is described. Thereafter, the methods for the inference are presented.
\subsection{Penalized Maximum Likelihood Estimation}
The resulting loglikelihood of the model is:
\begin{align*}
ll( \boldsymbol{\gamma} ) &= \sum_{n=1}^{N} \sum_{t_n=1}^{T_n}  \log \{ c(F_1(y_{nt_n1}),\hdots, F_M(y_{nt_nM}), \delta_{nt_n})  \} + \sum_{m=1}^{M} \left[  \log \{ f_m(y_{nt_nm}) \} \right]  \\
\end{align*}
adding the penalty results in the expression:
\begin{align} \label{llpen}
ll( \boldsymbol{\gamma})_{pen} = ll( \boldsymbol{\gamma}) - \frac{1}{2} \boldsymbol{\gamma}^T \boldsymbol{S} \boldsymbol{\gamma}
\end{align}
with  $\boldsymbol{\gamma}=(\boldsymbol{\beta}_1, \hdots, \boldsymbol{\beta}_M)$ and $\boldsymbol{S}=(\boldsymbol{D}_{\lambda 1},\hdots, \boldsymbol{D}_{\lambda M})$. \\ The model parameters $\boldsymbol{\gamma}$ and $\boldsymbol{\lambda}=(\lambda_1^\theta,\hdots,\lambda_{J^\theta_M}^\theta)$ can be estimated in the procedure described in \cite{marra2017simultaneous}. For the reader's convenience, it is summarized using this works notation. Let $\theta, \theta' \in \{\mu_1, \sigma_{V1}, \sigma_{U1}, \hdots,  \mu_M, \sigma_{VM}, \sigma_{UM}, \delta \}$ and $j^\theta_m \in \{0,1,\hdots,Q_{J^\theta_m} \}$. Introducing $i$ as an iteration variable. For a fixed parameter vector of the smoothing parameters $\boldsymbol{\lambda}^i$ defining the penalized gradient and hessian of iteration $i$ as $(\boldsymbol{g}_{pen})^i = (\boldsymbol{g})^i - \boldsymbol{S} (\boldsymbol{\gamma})^i$ and  $(\boldsymbol{H}_{pen})^i=(\boldsymbol{H})^i - \boldsymbol{S}$. The gradient and the hessian being  defined as follows:
\begin{align*}
 (\boldsymbol{g})^i &= ({(\boldsymbol{g}^{\mu_1})}^i, {(\boldsymbol{g}^{\sigma_{V1}})}^i, {(\boldsymbol{g}^{\sigma_{U1}})}^i,\hdots,{(\boldsymbol{g}^{\mu_M})}^i, {(\boldsymbol{g}^{\omega_M})}^i, {(\boldsymbol{g}^{\sigma_{UM}})}^i, {(\boldsymbol{g}^\delta)}^i ) \\
 (\boldsymbol{H})^i &= \begin{pmatrix}
 (\boldsymbol{H}^{\mu_1 \mu_1 })^i& {(\boldsymbol{H}^{\mu_1 \sigma_{V1} })}^i& {(\boldsymbol{H}^{\mu_1 \sigma_{U1}})}^i&\hdots & {(\boldsymbol{H}^{\mu_1 \delta})}^i\\
 {(\boldsymbol{H}^{\sigma_{V1} \mu_1 })}^i& {(\boldsymbol{H}^{\sigma_{V1} \sigma_{V1} })}^i& {(\boldsymbol{H}^{\sigma_{V1} \sigma_{U1}})}^i&\hdots & {(\boldsymbol{H}^{\sigma_{V1} \delta})}^i\\
 {(\boldsymbol{H}^{\sigma_{U1} \mu_1 })}^i& {(\boldsymbol{H}^{\sigma_{U1} \sigma_{V1} })}^i& {(\boldsymbol{H}^{\sigma_{U1} \sigma_{U1}})}^i&\hdots& {(\boldsymbol{H}^{\sigma_{U1} \delta})}^i& \\
 \vdots & & & \ddots & \vdots\\
 {(\boldsymbol{H}^{\delta \mu_1 })}^i& {(\boldsymbol{H}^{\delta \sigma_{V1} })}^i& {(\boldsymbol{H}^{\delta \sigma_{U1}})}^i&\hdots & {(\boldsymbol{H}^{\delta \delta})}^i\\
 \end{pmatrix} \\
 & \text{ where } \\
    {(\boldsymbol{g}^{\theta})}^i =& \frac{\partial ll( \boldsymbol{\gamma}) }{ \partial \boldsymbol{\beta}^{\theta}}|_{\boldsymbol{\beta}^{\theta}=(\boldsymbol{\beta}^{\theta})^i} \\
{(\boldsymbol{H}^{\theta,\theta'})}^i=& \frac{\partial^2 ll( \boldsymbol{\gamma}) }{ \partial \boldsymbol{\beta}^{\theta} \partial {\boldsymbol{\beta}^{\theta'}}^T}|_{\boldsymbol{\beta}^{\theta}={(\boldsymbol{\beta}^{\theta})}^i,\boldsymbol{\beta}^{\theta'}={(\boldsymbol{\beta}^{\theta'})}^i}
\end{align*}
which are calculated analytically to speed up the computation.  The trust region algorithm is applied for a fast an reliable estimation of  $\boldsymbol{\gamma}^{i}$. It was introduced by \cite{geyer2015trust} and solves
    \[
    \boldsymbol{\gamma}^{i+1}=\boldsymbol{\gamma}^{i} + \underset{r:||r|| \leq \boldsymbol{\gamma}^i}{\mathrm{argmin}} 
    v l_p (\boldsymbol{\Delta}^{i} )
    \] 
    with  $v l_p (\boldsymbol{\gamma}^{i} )=-(ll( \boldsymbol{\gamma})_{pen}  + \boldsymbol{r}^T \boldsymbol{g}_{pen}^i + \frac{1}{2} \boldsymbol{r}^T \boldsymbol{H}_{pen}^i \boldsymbol{r} )$. Here $||\cdot||$ denotes the Euclidean norm. Further, $\Delta^i$ is the radius of the trust region, which is adjusted within the algorithm\footnote{For more details see \cite{geyer2015trust}}. Utilizing a trust region algorithm proved to be faster compared to line-search counterparts. The reasons are given by \cite{marra2017simultaneous} and \cite{radice2016copula}. \\
     In the second step of the estimation, the smoothing parameter $\boldsymbol{\lambda}^i$ is selected. One can rewrite the parameter estimator as
\begin{equation} \label{gammarewr}
    \boldsymbol{\gamma}^{i+1}=(-\boldsymbol{H}^i + \boldsymbol{S})^{-1} \sqrt{-\boldsymbol{H}^i} \boldsymbol{B}^{i}
\end{equation}
where $\boldsymbol{B}^{i} = \sqrt{-\boldsymbol{H}^i} \boldsymbol{\gamma}^{i} + \boldsymbol{e}^{i}$ and $\boldsymbol{e}^{i} = \sqrt{-\boldsymbol{H}^i}^{-1} \boldsymbol{g}^{i}$. Maximum likelihood theory states that $\boldsymbol{e} \sim N(\boldsymbol{0}, \boldsymbol{I})$ and $\boldsymbol{B} \sim N(\sqrt{-\boldsymbol{H}} \boldsymbol{\gamma}^0, \boldsymbol{I})$  where $\boldsymbol{\gamma}^0$
denotes the true parameter vector. Consequently, for some estimate $\hat{\boldsymbol{\gamma}}$ one gets the expected value vector for $\boldsymbol{B}$  given by $\sqrt{-\boldsymbol{H}} \hat{\boldsymbol{\gamma}} = \boldsymbol{A} \boldsymbol{B}$, where $\boldsymbol{A}=\sqrt{-\boldsymbol{H}}(-\boldsymbol{H}+\boldsymbol{S})^{-1} \sqrt{-\boldsymbol{H}}$. Representation (\ref{gammarewr}) allows us to base smoothing parameter estimation on a parameterization of $\boldsymbol{B}$ that uses $\boldsymbol{H}$ and $\boldsymbol{g}$ as
a whole instead of the $\sum_{n=1}^N T_n$ components that make them up. The smoothing
parameter vector is estimated so that $\sqrt{-\boldsymbol{H}} \hat{\boldsymbol{\gamma}}$ is as close as possible to $\sqrt{-\boldsymbol{H}} \boldsymbol{\gamma}^0$. Using this concept, for a given $\boldsymbol{\gamma}^{i+1}$, the problem becomes
\begin{align*}
\boldsymbol{\lambda}^{i+1} =& \underset{\boldsymbol{\lambda}}{arg \min} ||\boldsymbol{B}^{i+1} - \boldsymbol{A}^{i+1} \boldsymbol{B}^{i+1}||^2 \\
&-(|\{\mu_1, \sigma_{V1}, \sigma_{U1}, \hdots,  \mu_M, \sigma_{V1}, \sigma_{UM}, \delta \}| \sum_{n=1}^N T_n) + 2 \cdot trace(\boldsymbol{A}^{i+1}) \qquad,
\end{align*} which can be solved by the algorithm introduced by \cite{wood2004stable}. If no non-linear or random effects are modelled, the second step is obsolete.

\subsection{Model Choice and Inference}
\cite{wojtys2015copula} prove the consistency of the estimator previously introduced, which is in principle applicable for any model fitted by penalized maximum likelihood, see \cite{marra2017simultaneous}. To obtain confidence intervals of the model coefficients  in the \textit{GJRM} one may utilize the approach of \cite{wood2017generalized}. 
The idea is to rewrite the model into a fully Bayesian model and then simulate from the resulting posterior distribution. If one takes a Bayesian view on the smoothing process in which $\boldsymbol{\gamma}$ has a zero mean improper Gaussian prior distribution with precision matrix proportional to $\boldsymbol{S}$ then $\boldsymbol{\gamma}|\boldsymbol{y},\boldsymbol{\lambda} \sim N(\hat{\boldsymbol{\gamma}},\hat{\boldsymbol{H}}_{pen}^{-1})$.
Deduction of Bayesian credible intervals remains a trivial task. \cite{wood2017generalized} showed that the Bayesian credible intervals for a component smooth function are close to the observed coverage of a frequentist confidence interval. 


When working with real world data, the data generating process and thus the true copula is unknown. As a result, the researcher needs to make a reasoned choice among a multitude of copula models. Model choice criteria like Generalized Akaike Information Criterion (GAIC) can be applied here. 
The GAIC was introduced by \cite{akaike1983information} and is defined as: 
\[
GAIC=-2 ll(\hat{\boldsymbol{\gamma}}) + 2 df \qquad,
\]
here $df$ denotes the total effective degrees of freedom of the model. When smoothers are applied, the $df$ do not necessarily correspond with the number of parameters of the model. Instead
\[
df= \sum_{\theta } trace (\boldsymbol{Z}^{\theta}{({\boldsymbol{Z}^\theta}^T \boldsymbol{Z}^\theta+ \boldsymbol{D}^\theta)}^{-1} {\boldsymbol{Z}^\theta}^T) \qquad.
\]
The GAIC is negatively oriented, thus the model with the smallest value is then chosen.

\section{Monte Carlo Simulation Study} \label{MC simulation}
To demonstrate the practicality of the suggested approach, a Monte Carlo simulation experiment was conducted. The underlying data generating process (dgp) is specified in the following section.

\subsection{Monte Carlo Scenario Settings}
The number of DMUs is $N \in \{50, 100, 150\}$. To create an unbalanced panel data set, each observation is assigned to a DMU utilizing a discrete uniform distribution with support $\{1,2, \hdots N\}$. The total number of observations is set to $\sum_n^N T_n = 10 N$. The number of Montecarlo Simulation (NMC) is $1000$, to resemble large sample properties. The independent variables \\
$x^{\mu}_{nt_nm}, x^{\sigma}_{Vnt_nm}, x^{\alpha}_{nt_nm}, x^{\delta}_{nt_n} \sim U(-1,1)$, while the random effects $re_{n} \sim N(0,0.25)$. The number of sub-DMUs is set to 2, e.g. $M=2$. Further, $s_1=-1$ and $s_2=1$. Further, $U_{nt_n1} \sim HN(\sigma_{Unt_n1}^2)$ and $U_{nt_n2} \sim HN(\sigma_{Unt_n2})$. All introduced splines are of degree $3$ with $10$ evenly spaced knots. 
\begin{align*}
    \eta_{nt_n1}^{\mu}(\boldsymbol{x}_{nt_n1}^{\mu})  =& \pi^2 \tanh(\pi (x^{\mu}_{nt_n1} +1) \frac{4}{5}) + re_n \\
    \eta_{nt_n1}^{\sigma_V}(\boldsymbol{x}_{nt_n1}^{\sigma_V}) =&-1.5+0.75 x^{\sigma_V}_{nt_n1}  \\
    \eta_{nt_n1}^{\sigma_U}(\boldsymbol{x}_{nt_n1}^{\sigma_U}) =& -1.4+0.6 x^{\sigma_U}_{nt_n1}\\
    \eta_{nt_n2}^{\mu}(x_{nt_n2}^{\mu})  =& 6 -3.5 x^{\mu}_{nt_n2}+1.75 (x^{\mu}_{nt_n2})^2 + re_n \\
    \eta_{nt_n2}^{\sigma_V}(\boldsymbol{x}_{nt_n2}^{\sigma_V}) =&  -1.5+0.75x^{\sigma_V}_{nt_nm}\\
    \eta_{nt_n2}^{\sigma_U}(\boldsymbol{x}_{nt_n2}^{\sigma_U}) =& 1.25+2x^{\sigma_U}_{nt_n2}   \\
    \eta_{nt_n}^{\delta}(\boldsymbol{x}_{nt_n}^{\delta}) =& 1.2+2.5 x^{\delta}_{nt_n} + re_n \\
\end{align*}
To measure the quality of the estimated parameters, the Mean Integrated Squared Error (MISE) is utilized. It's estimate is defined as follows
\[
MISE(\theta) = \frac{1}{NMC} \sum_{i=1}^{NMC}  \left[ \int_{-1}^{1} (\widehat{\eta_{i}}^{\theta}(\boldsymbol{x}_{i}^{\theta})-\eta^{\theta}(\boldsymbol{x}_{i}^{\theta}))^2  dx^\theta\right] \qquad,
\]
where $\widehat{\eta_{i}}^{\theta}(\boldsymbol{x}_{i}^{\theta})$ is the functional estimate of $\eta^{\theta}(\boldsymbol{x}^{\theta}_i)$ obtained in the respective simulation $i \in \{1,2,\hdots,NMC\}$.
\subsection{Monte Carlo Scenario Results}
In the Table \ref{tab:MISE} the \textit{MISE} for each parameter of the model is reported. \\
\begin{table}[H]
\begin{tabular}{ c c c c c c c c c c} 
\hline
N &  $\mu_1$ & $\sigma_{V1}$ & $\sigma_{U1}$ &  $\mu_2$ & $\sigma_{V2}$ & $\sigma_{U2}$ & $\delta$ \\
\hline
  50 & 0.0029000 & 0.0156483 & 0.0353771 & 0.0023770 & 0.0149879 & 0.0259144 & 0.0953214 \\  
  100 & 0.0015482 & 0.0058820 & 0.0208352 & 0.0012734 & 0.0058574 & 0.0142832 & 0.0423325 \\
  150 & 0.0010267 & 0.0034294 & 0.0148707 & 0.0008836 & 0.0037562 & 0.0112187 & 0.0242430 \\
\hline  
\end{tabular}
\caption{MISE for the simulation scenario for all $N$ and distribution parameters}
\label{tab:MISE}
\end{table}
It shows, that if $N$ increases that the MISE decreases, which coincides with ML theory.
Further, the plots of the estimated and true functions of $\eta^{\mu_1}, \eta^{\sigma_{U1}}, \eta^{\mu_2},$ and $\eta^{\sigma_{U2}}$ for $N=150$ are provided.
\begin{figure}[H]
     \centering
     \begin{subfigure}{1\textwidth}
         \centering
         \includegraphics[width=0.45\textwidth]{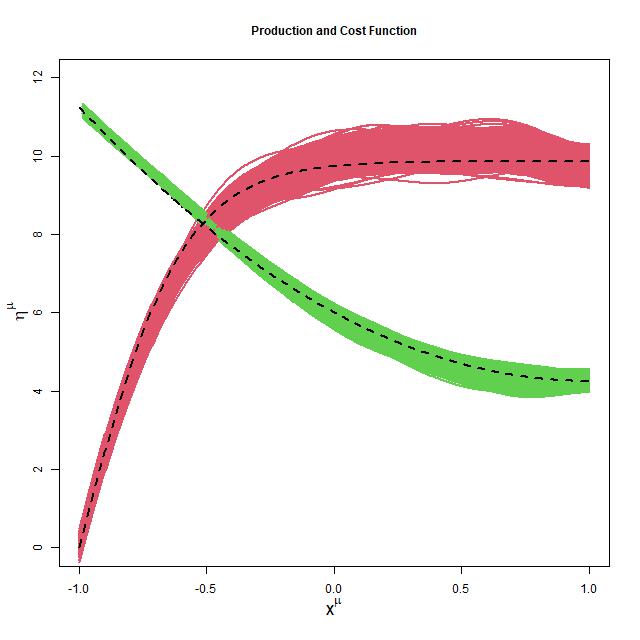} 
         \includegraphics[width=0.45\textwidth]{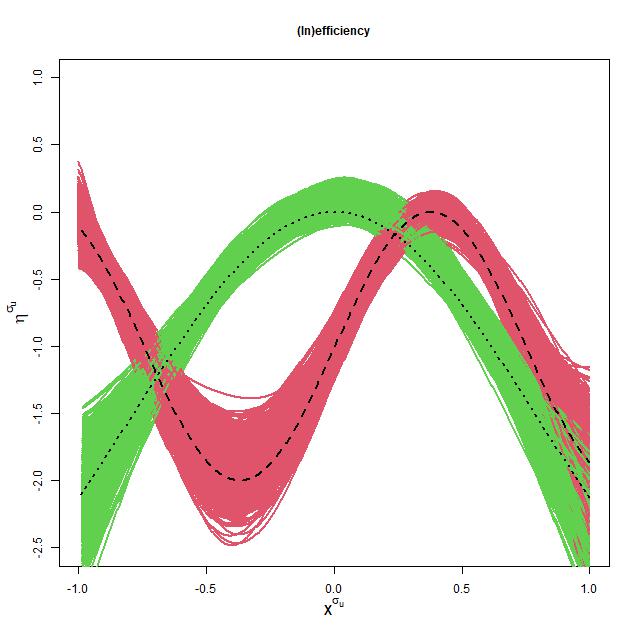}
     \end{subfigure}
        \caption{Plot of the estimated effect of $x_{nt_nm}^\theta$ on $h(x_{nt_nm}^\theta)$.  For $m=1$, the dashed black line is the true function, while the red solid lines are the estimated ones. For $m=2$, the dotted black line is the true function, while the green solid lines are the estimated ones.}
\end{figure}
Thus, the simulation study has shown that the performance of the estimator is satisfactory and improves as the number of observations increases in all three scenarios with excellent accuracy for the approximation of the production function.

\newpage

\section{Conclusions} \label{conclusions}
In this work, the modeling framework for the \textit{Distributional Stochastic Frontier Model} was introduced. The model allows for a high degree of flexibility by allowing for the influence of covariates on all parameters of the distribution of the output and exploiting the strength of penalized B-splines. The model can be utilized for cross-sectional and unbalanced panel data. By relaxing the strong assumption of a parametric production (cost) function, the model is less likely to lead to false conclusions. Further, it allows capturing the potentially non-linear dependence between multiple outputs. Due to the modularity of the estimation approach, other distributional assumptions about the composed error terms can be included easily. The parameters of the joint composed error terms are estimated via penalized maximum likelihood estimation with hyperparameter choice via GAIC. Confidence intervals and significance test can be deducted from the estimation results. The model has been demonstrated using simulated data.

\begin{acknowledgements}
The authors would like to thank Alexander Ritz for his insights and helpful comments. Financial support from the Deutsche Forschungsgemeinschaft (DFG, German Research Foundation) grant KN 922/9-1 is gratefully acknowledged. 
\end{acknowledgements}

%
\section*{Conflict of interest}
The authors declare that they have no conflict of interest.

\newpage

\section{Appendix}

\subsection{P-splines} \label{P-splines}
Here, a brief introduction in penalized B-splines terms is given. For more details, see \cite{eilers1996flexible}. For the sake of simplicity, the indexes $n, m, \theta$ and $j$ are dropped in this section, where not necessary. The function of interest, $h(\cdot)$ can be approximated by piecewise polynomials. A spline of order $l+1$ is a piecewise polynomial function of degree $l$ in the variable $x$. The values of $x$ where the pieces of polynomial meet are known as knots $\kappa_1, \kappa_2,\ldots,\kappa_{n'}$. These are unique and in an increasing order. The knots may be equidistant or based on quantiles. B-splines of order $l+1$ are basis functions for spline functions of the same order defined over the same knots. Consequently, all possible spline functions can be set up from a linear combination of B-splines. Further, the linear combination for each spline function is unique. Thus, if the complete basis is utilized, the function $h(x)$ can be represented through a linear combination of $n'+l-1=Q$ basis functions:
\[
h(x) = \sum_{q}^{Q}  b^l_{q}(x) \beta_{q} \qquad.
\]
The definition for B-splines of order $l=0$ is given by:
\[
b^0_{q}(x)=I(\kappa_q \leq x \leq \kappa_{q+1}) \qquad q=1,\ldots,Q+1 \qquad.
\]
Higher order B-splines are defined recursively:
\[
b^l_{q}(x)=\frac{x-\kappa_{q-l}}{\kappa_{q}-\kappa_{q-l}}b^{l-1}_{q-1}(x) +\frac{\kappa_{q+1} - x}{\kappa_{q+1}-\kappa_{q+1-l}}b^{l-1}_{q}(x) \qquad.
\]
The number and position of the knots influences the fit of the function.  Thus, they should be chosen carefully. \cite{eilers1996flexible} propose to use a relatively large number of knots and introduce a penalty for functions that are not smooth enough. The smoothness can be approximated by a difference of coefficients of adjacent B-splines:
\[
pen_{smooth}(\boldsymbol{\beta}, \lambda) = \frac{1}{2} \lambda \sum_{q+2}^{Q} (\Delta^2 (\beta_{q}))^2  =  \frac{1}{2} \lambda \boldsymbol{\beta}^T \underbrace{\boldsymbol{P}^T \boldsymbol{P}}_{\boldsymbol{D}} \boldsymbol{\beta}  \qquad,
\]
where $\Delta^d (\beta_{q})$ is the $d$-th order difference operator of the vector $\boldsymbol{\beta}=(\beta_{1},\hdots,\beta_{Q})$. Here $\boldsymbol{D}$ is the difference matrix. The parameter $\lambda> 0$ influences the smoothness of the fitted curve. For small values of $\lambda$ the estimated function will be closer to the data, for high values of $\lambda$ the estimated curve will be smoother. For $\lambda \to \infty$ the curve will approximate a polynomial of degree $d-1$. Thus the degrees of freedom are influenced by $\lambda$. The parameter $\lambda$ can be found using the GAIC.

\newpage

\subsection{SC-Splines} \label{SC-splines}
If one wants to enforce monotonically increasing and concavity shape constraints, the $\boldsymbol{\Sigma}$ is build in the following way:
 \begin{align*} 
     \boldsymbol{\Sigma}[i,k]_j=\begin{cases}
     0  \hfill &\text{ if } i = 1, k \geq 2 \\
     1 \hfill &\text{ if } i \geq 1, k = 1 \\
     i-1 \hfill &\text{ if } i \geq 2, k = 2,3,\hdots,Q_j-1+2 \\
     Q_j-k+1 \hfill &\text{ if } i \geq 2, k = Q_j-i+3, Q_j-i+2,\hdots,Q_j \\
     \end{cases}
\end{align*}
while for no constraints $\boldsymbol{\Sigma}_j=\boldsymbol{I}_j$. The quadratic penalty $\boldsymbol{D}_j$ is set up
\begin{align*} 
     \boldsymbol{D}[i,k]_j=\begin{cases}
     1 \hfill &\text{ if } i = 1,\hdots,Q_j-3, k = i+2 \\
     -1 \hfill &\text{ if }i = 1,\hdots,Q_j-3, k = i+3 \\
     0 \hfill &\text{ otherwise }   \quad.\\
     \end{cases}
\end{align*} For other constraints, see the previously mentioned paper which defines constraint specific $\boldsymbol{\Sigma}_j$ and $\boldsymbol{D}_j$.
 
\subsection{Copula Densities} \label{copulapdf}
In the following section to density and inverse link functions of the Normal, Clayton, and Gumbel copula are provided. They are required for the evaluation of the log likelihood of the model in the simulation study.
    \begin{align*}
        c_{normal}(w_1,w_2,\delta)=&-\frac{1}{2} \log \{1-\delta^2\} + \frac{\delta}{1-\delta^2} \Phi^{-1}(w_1) \Phi^{-1}(w_2)- \frac{\delta^2}{2(1-\delta^2)}(\Phi^{-1}(w_1)^2 +\Phi^{-1}(w_2)^2) \\
        g^{-1}_{normal}(\eta)=&\frac{\eta}{\sqrt{1+\eta^2}} \\
     c_{clayton}(w_1,w_2,\delta)=&\frac{(1+\delta)(w_1^{-\delta}+w_2^{-\delta}-1)^{-\frac{1}{\delta}-2}}{(w_1 w_2)^{\delta+1}} \\
        g^{-1}_{clayton}(\eta)=&\exp\{ \eta\} \\
     c_{gumbel}(w_1,w_2,\delta)=& \frac{\exp\{-((-\log \{ w_1 \})^\delta+(-\log \{w_2 \})^\delta)^{\frac{1}{\delta}} \}}{w_1 w_2 ((-\log \{ w_1 \})^\delta + (-\log\{w_2 \})^\delta)^{2-\frac{1}{\delta}}}\\
     & \times (\log \{ w_1 \} \log \{ w_2 \})^{\delta-1} (((-\log \{w_1 \})^\delta + (- \log \{ w_2\})^\delta)^\frac{1}{\delta}+\delta-1\\
        g_{gumbel}^{-1}(\eta)=&\exp\{ \eta\} +1 
    \end{align*}

\newpage


\newpage


\bibliographystyle{spbasic}      
\bibliography{citations}   

\begin{thebibliography}{28}
\providecommand{\natexlab}[1]{#1}
\providecommand{\url}[1]{{#1}}
\providecommand{\urlprefix}{URL }
\expandafter\ifx\csname urlstyle\endcsname\relax
  \providecommand{\doi}[1]{DOI~\discretionary{}{}{}#1}\else
  \providecommand{\doi}{DOI~\discretionary{}{}{}\begingroup
  \urlstyle{rm}\Url}\fi
\providecommand{\eprint}[2][]{\url{#2}}

\bibitem[{Aigner et~al.(1977)Aigner, Lovell, and
  Schmidt}]{aigner1977formulation}
Aigner D, Lovell CK, Schmidt P (1977) Formulation and estimation of stochastic
  frontier production function models. Journal of econometrics 6(1):21--37

\bibitem[{Akaike(1983)}]{akaike1983information}
Akaike H (1983) Information measures and model selection. Int Stat Inst
  44:277--291

\bibitem[{Arellano-Valle and Azzalini(2006)}]{arellano2006unification}
Arellano-Valle RB, Azzalini A (2006) On the unification of families of
  skew-normal distributions. Scandinavian Journal of Statistics 33(3):561--574

\bibitem[{Battese and Coelli(1988)}]{battese1988prediction}
Battese GE, Coelli TJ (1988) Prediction of firm-level technical efficiencies
  with a generalized frontier production function and panel data. Journal of
  econometrics 38(3):387--399

\bibitem[{Chambers et~al.(1988)}]{chambers1988applied}
Chambers RG, et~al. (1988) Applied production analysis: a dual approach.
  Cambridge University Press

\bibitem[{Dom{\'\i}nguez-Molina et~al.(2007)Dom{\'\i}nguez-Molina,
  Gonz{\'a}lez-Far{\'\i}as, Ramos-Quiroga, and Gupta}]{dominguez2007matrix}
Dom{\'\i}nguez-Molina JA, Gonz{\'a}lez-Far{\'\i}as G, Ramos-Quiroga R, Gupta AK
  (2007) A matrix variate closed skew-normal distribution with applications to
  stochastic frontier analysis. Communications in Statistics—Theory and
  Methods 36(9):1691--1703

\bibitem[{Eilers et~al.(1996)Eilers, Marx et~al.}]{eilers1996flexible}
Eilers PH, Marx BD, et~al. (1996) Flexible smoothing with b-splines and
  penalties. Statistical science 11(2):89--121

\bibitem[{Fan et~al.(1996)Fan, Li, and Weersink}]{fan1996semiparametric}
Fan Y, Li Q, Weersink A (1996) Semiparametric estimation of stochastic
  production frontier models. Journal of Business \& Economic Statistics
  14(4):460--468

\bibitem[{Ferrara and Vidoli(2017)}]{ferrara2017semiparametric}
Ferrara G, Vidoli F (2017) Semiparametric stochastic frontier models: A
  generalized additive model approach. European Journal of Operational Research
  258(2):761--777

\bibitem[{Geyer(2015)}]{geyer2015trust}
Geyer CJ (2015) Trust region optimization. No Journal

\bibitem[{Giannakas et~al.(2003)Giannakas, Tran, and
  Tzouvelekas}]{giannakas2003choice}
Giannakas K, Tran KC, Tzouvelekas V (2003) On the choice of functional form in
  stochastic frontier modeling. Empirical Economics 28(1):75--100

\bibitem[{Jondrow et~al.(1982)Jondrow, Lovell, Materov, and
  Schmidt}]{jondrow1982estimation}
Jondrow J, Lovell CK, Materov IS, Schmidt P (1982) On the estimation of
  technical inefficiency in the stochastic frontier production function model.
  Journal of econometrics 19(2-3):233--238

\bibitem[{Klein et~al.(2020)Klein, Herwartz, and Kneib}]{klein2020modelling}
Klein N, Herwartz H, Kneib T (2020) Modelling regional patterns of
  inefficiency: A bayesian approach to geoadditive panel stochastic frontier
  analysis with an application to cereal production in england and wales.
  Journal of Econometrics 214(2):513--539

\bibitem[{Kumbhakar et~al.(2007)Kumbhakar, Park, Simar, and
  Tsionas}]{kumbhakar2007nonparametric}
Kumbhakar SC, Park BU, Simar L, Tsionas EG (2007) Nonparametric stochastic
  frontiers: a local maximum likelihood approach. Journal of Econometrics
  137(1):1--27

\bibitem[{Kuosmanen(2008)}]{kuosmanen2008representation}
Kuosmanen T (2008) Representation theorem for convex nonparametric least
  squares. The Econometrics Journal 11(2):308--325

\bibitem[{Kuosmanen and Johnson(2010)}]{kuosmanen2010data}
Kuosmanen T, Johnson AL (2010) Data envelopment analysis as nonparametric
  least-squares regression. Operations Research 58(1):149--160

\bibitem[{Lai(2020)}]{lai2020maximum}
Lai Hp (2020) Maximum simulated likelihood estimation of the seemingly
  unrelated stochastic frontier regressions. Empirical Economics pp 1--26

\bibitem[{Lai and Huang(2013)}]{lai2013maximum}
Lai Hp, Huang CJ (2013) Maximum likelihood estimation of seemingly unrelated
  stochastic frontier regressions. Journal of Productivity Analysis 40(1):1--14

\bibitem[{Marra and Radice(2017)}]{marra2017bivariate}
Marra G, Radice R (2017) Bivariate copula additive models for location, scale
  and shape. Computational Statistics \& Data Analysis 112:99--113

\bibitem[{Marra et~al.(2017)Marra, Radice, B{\"a}rnighausen, Wood, and
  McGovern}]{marra2017simultaneous}
Marra G, Radice R, B{\"a}rnighausen T, Wood SN, McGovern ME (2017) A
  simultaneous equation approach to estimating hiv prevalence with nonignorable
  missing responses. Journal of the American Statistical Association
  112(518):484--496

\bibitem[{Meeusen and van Den~Broeck(1977)}]{meeusen1977efficiency}
Meeusen W, van Den~Broeck J (1977) Efficiency estimation from cobb-douglas
  production functions with composed error. International economic review pp
  435--444

\bibitem[{Pya and Wood(2015)}]{pya2015shape}
Pya N, Wood SN (2015) Shape constrained additive models. Statistics and
  Computing 25(3):543--559

\bibitem[{Radice et~al.(2016)Radice, Marra, and Wojty{\'s}}]{radice2016copula}
Radice R, Marra G, Wojty{\'s} M (2016) Copula regression spline models for
  binary outcomes. Statistics and Computing 26(5):981--995

\bibitem[{Ruppert et~al.(2003)Ruppert, Wand, and
  Carroll}]{ruppert2003semiparametric}
Ruppert D, Wand MP, Carroll RJ (2003) Semiparametric regression. 12, Cambridge
  university press

\bibitem[{Stasinopoulos et~al.(2007)Stasinopoulos, Rigby
  et~al.}]{stasinopoulos2007generalized}
Stasinopoulos DM, Rigby RA, et~al. (2007) Generalized additive models for
  location scale and shape (gamlss) in r. Journal of Statistical Software
  23(7):1--46

\bibitem[{Wojty{\'s} and Marra(2015)}]{wojtys2015copula}
Wojty{\'s} M, Marra G (2015) Copula based generalized additive models with
  non-random sample selection. arXiv preprint arXiv:150804070

\bibitem[{Wood(2004)}]{wood2004stable}
Wood SN (2004) Stable and efficient multiple smoothing parameter estimation for
  generalized additive models. Journal of the American Statistical Association
  99(467):673--686

\bibitem[{Wood(2017)}]{wood2017generalized}
Wood SN (2017) Generalized additive models: an introduction with R. CRC press

\end{thebibliography}

\end{document}